 \documentclass[final,authoryear,12pt]{elsarticle}

 \usepackage{graphicx}

\usepackage{amssymb}

\journal{Planetary and Space Science}

\begin{document}

\begin{frontmatter}





\title{The Cratering History of Asteroid (2867) Steins}

%
\author[mar]{S. Marchi\corref{cor1}}
\ead{simone.marchi@unipd.it}
\author[mar]{C. Barbieri}
\author[kue]{M. K\"uppers} 
\author[marz]{F. Marzari}  
\author[dav]{B. Davidsson}
\author[kel]{H.U. Keller}
\author[bes]{S. Besse}
\author[bes]{P. Lamy}
\author[mot]{S. Mottola}    
\author[mas]{M. Massironi}
\author[cre]{G. Cremonese}
%
\cortext[cor1]{Corresponding author}
\address[mar]{Department of Astronomy, Padova University, Italy}
\address[kue]{ESA-ESAC, Villanueva de la Ca\~nada (Madrid), Spain}
\address[marz]{Department of Physics, Padova University, Italy}
\address[dav]{Department of Physics and Astronomy, Uppsala University, Sweden}
\address[kel]{Max Planck Institute for Solar System Research, Lindau, Germany}
\address[bes]{Laboratoire d'Astrophysique de Marseille, France}
\address[mot]{DLR-Berlin, Germany}
\address[mas]{Department of Geosciences, Padova University, Italy}
\address[cre]{INAF-Padova, Italy}

\begin{abstract}

The cratering  history of  main belt asteroid  (2867) Steins  has been
investigated using  OSIRIS imagery  acquired during the  Rosetta flyby
that took place on the  5$^{th}$ of September 2008.  For this purpose,
we applied  current models describing  the formation and  evolution of
main belt asteroids, that  provide the rate and velocity distributions
of  impactors. These  models coupled  with appropriate  crater scaling
laws, allow the  cratering history to be estimated.   Hence, we derive
Steins'  cratering retention  age, namely  the time  lapsed  since its
formation or global surface  reset.  We also investigate the influence
of various  factors -like  bulk structure and  crater erasing-  on the
estimated age, which spans from a few hundred Myrs to more than 1~Gyr,
depending on the adopted scaling law and asteroid physical parameters.
Moreover, a  marked lack  of craters smaller  than about  $0.6$~km has
been  found and interpreted  as a  result of  a peculiar  evolution of
Steins cratering  record, possibly related either to  the formation of
the  2.1~km  wide  impact  crater  near  the south  pole  or  to  YORP
reshaping.

\end{abstract}


\end{frontmatter}



\section{Introduction}

The  European Space Agency's  (ESA) Rosetta  spacecraft passed  by the
main belt asteroid (2867) Steins  with a relative velocity of 8.6~km/s
on 5 September  2008 at 18:38:20 UTC.  The  Rosetta-Steins distance at
closest approach  (CA) was 803~km.   During the flyby the  solar phase
angle (sun-object-observer) decreased from the initial 38$^{\circ}$ to
a minimum of 0.27$^{\circ}$ two  minutes before CA and increased again
to 51$^{\circ}$  at CA, to  reach 141$^{\circ}$ when  the observations
were stopped.  A  total of 551 images were  obtained by the scientific
camera system  OSIRIS, which consists  of two imagers: the  Wide Angle
Camera (WAC) and  the Narrow Angle Camera (NAC)  \citep{kel07}.
The best  resolution at CA corresponded  to a scale of  80~m/px at the
asteroid surface.\\
Steins  has   an  orbital  semi-major   axis  of  about   2.36~AU,  an
eccentricity  of 0.15  and  an inclination  of  9.9$^{\circ}$.  It  is
therefore orbiting in a relatively  quiet region of the main belt, far
from the  main escape  gateways, namely the  secular $\nu_6$  and mean
motion 3:1 resonances.   Its shape can be  fitted by an ellipsoid
  having axis of $6.67 \times 5.81 \times 4.47$~km \citep{kel10}.\\
Previous space missions have visited  and acquired detailed data for a
total  of 5  asteroids, namely  three main  belt  asteroids \citep[951
  Gaspra,  243  Ida,   253  Mathilde;][]{vev99a,bel92,bel94}  and  two
near-Earth  objects \citep[433 Eros,  25143 Itokawa;][]{vev99b,sai06}.
Itokawa is the smallest of  them, with dimensions of $0.45 \times 0.29
\times 0.21$~km.  The  other asteroids have average sizes ranging
  from about 12~km to 53~km.  In this respect, Steins with its 5.3~km
size lies between Itokawa and the large asteroids. It is therefore the
smallest main  belt asteroid ever  imaged by a spacecraft  (except for
Ida's satellite  Dactyl with  a diameter of  roughly a  km). Moreover,
Steins is  a member of the  relatively rare E-type  class (composed by
igneous materials),  while other  asteroids visited by  spacecraft are
members of the most common S- and C-type classes.  Previous spacecraft
observations opened a new field of investigation, namely the cratering
of  asteroidal  surfaces.   A  number of  interesting  processes  were
therefore studied with unprecedented detail, like the cratering on low
gravity     bodies,     regolith     formation,    seismic     shaking
\citep[e.g.][]{cha02}.\\
%
This paper analyzes some of  the highest resolution OSIRIS images with
the  aim  to  study  the  crater  size  distribution  and  derive  the
chronology of the  impacts on the surface of  the asteroid.  This will
also provide clues  on the Steins bulk structure,  evolution, and give
new insights on the above mentioned processes.\\

\section{Steins crater population and geological assessment}

\begin{figure*}[h]
\includegraphics[width=6cm]{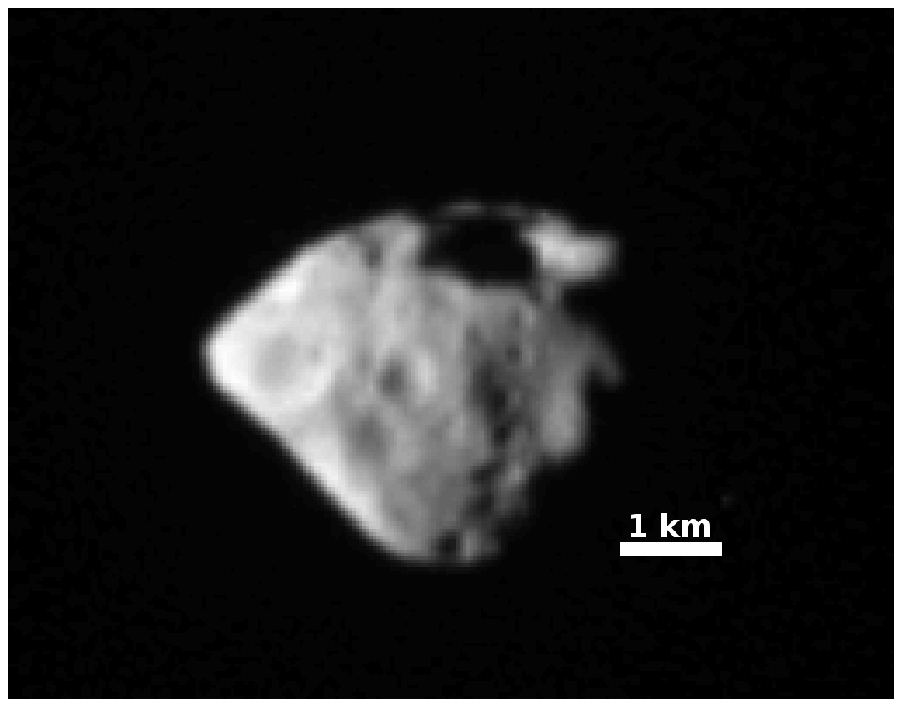}
\includegraphics[width=6cm]{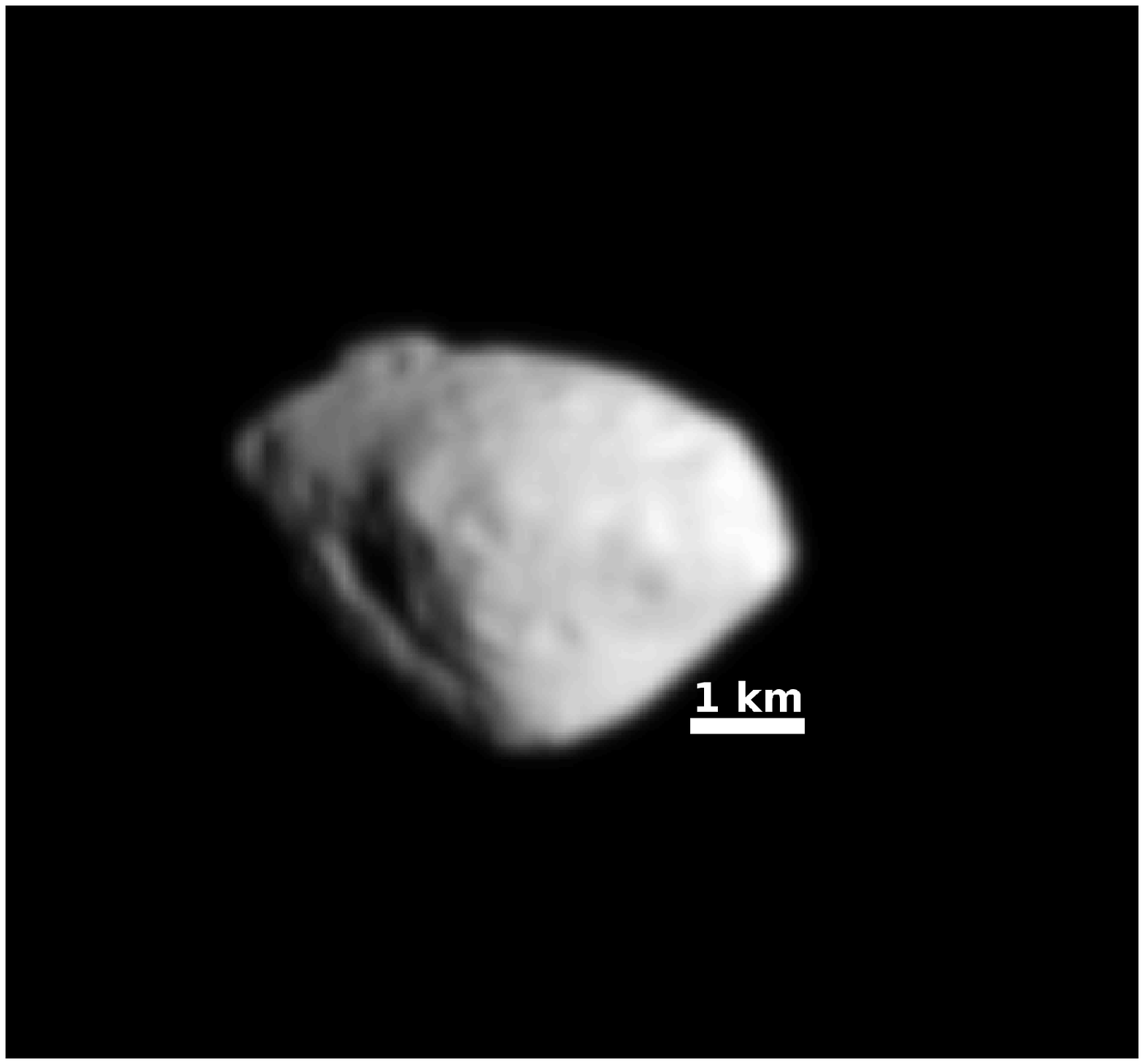}
\caption{WAC  image  (left) at  the  close  approach.   The image  was
  acquired with  a narrow passband  filter with central  wavelength of
  295.9~nm  and  FWHM=10.9~nm.   The  spatial resolution  is  80~m/px.
  Since  craters close  to the  limb  are not  easily detectable,  the
  actual region used  for crater count is reduced  by one pixel around
  the limb.  The pit-chain is  clearly visible close to the right-hand
  terminator.  The NAC image  (right) shows the large depression which
  is approximately in the antipodal  position of the alignment of pits
  and  Ruby crater.   The image  was acquired  with a  narrow passband
  filter with  central wavelength  of 805.3~nm and  FWHM=84.5~nm.  The
  spatial resolution is 98~m/px.}
\label{wac-nac}
\end{figure*}

\begin{figure*}[h]
\includegraphics[width=11cm,angle=-90]{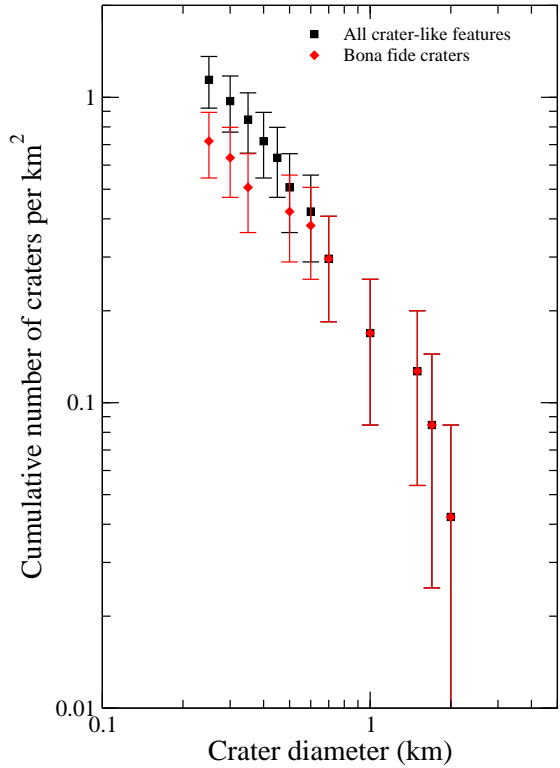}
\caption{Cumulative distribution of  all crater-like features and bona
  fide craters.   The latter is the  distribution used in  the rest of
  the paper  for age assessment.  Both  distributions contain features
  $\ge3$~pixels \citep[see][for more details]{bes10}.  Notice that the
  bona  fide craters  distribution  does not  contain the  depressions
  forming the chain-like features (see  text for details) and a couple
  of  uncertain  craters  close  to  the Ruby's  rim.   The  resulting
  distribution  shows a remarkable  paucity of  small ($D<0.5-0.6$~km)
  craters  when   compared  to   the  distribution  of   all  detected
  crater-like  features. Error  bars  are estimated  on  the basis  of
  Poisson statistics of counts.  }
\label{allcraters}
\end{figure*}

A total of 42 crater-like  features with dimensions ranging from about
0.2 to 2.1~km  have been identified on three  WAC images obtained near
closest approach  and one  NAC image obtained  about 10~min  before CA
(see fig. \ref{wac-nac}). However,  the illumination conditions of the
WAC   images   were  considerably   different   from   those  of   the
NAC. Therefore  in the present work,  we restrict the  analysis to the
WAC  images only.   The effective  area  over which  counts have  been
performed is 23.7~km$^2$, or approximately 25\% of the estimated total
Steins' surface of 97.6~km$^2$ \citep[for more details on crater
    identification and size estimate see][]{bes10}.\\
A total of 29 crater-like  features larger than 3 pixels, ranging from
about 0.24~km to 2.1~km, have been identified and used for the present
work.  The largest  crater (named Ruby hereinafter) has  a diameter of
2.1~km.   The   true  nature  of  some  features   is  uncertain.   In
particular,  this is  the  case for  a  feature consisting  of 8  pits
aligned along  a straight chain  crossing a large degraded  crater and
extending almost  from the  south to the  north pole of  the asteroid.
Owing to lack of sufficient resolution, it is not clear whether or not
these aligned  pits are  due to impacts.   The probability for  such a
chain of impacts to occur on a low gravity body like Steins is indeed
extremely low.  Moreover, a  NAC image (see fig.  \ref{wac-nac} bottom
panel),  obtained from  a different  aspect, shows  the presence  of a
large depression and an array of rimless circular pits, continuing the
pit-chain imaged  by the WAC \citep{kel10}.  Therefore,  the series of
aligned depressions are most  likely not caused by individual impacts.
An alternative explanation is that  the impact forming the crater Ruby
triggered the nucleation of a long fracture or fault, whose expression
at the  surface is the  pit-chain.  Similar linear  features were
  also  observed on  other small  bodies, like  Gaspra \citep{bel92}.
Other uncertain crater-like  features are located close to  the rim of
the Ruby crater \citep{bes10}.\\
The  cumulative
distributions of  all detected crater-like features and  18 bona
fide craters  are reported in fig.  \ref{allcraters}.   The latter has
been constructed rejecting all  the uncertain craters described above.
Note that  the small bona  fide craters ($D<0.5-0.6$~km)  are strongly
underrepresented with  respect to the distribution  of the crater-like
features.\\
The  population of  craters shows  a wide  range  of depth-to-diameter
ratios,  varying  from very  shallow  craters  ($\sim  0.05$) to  deep
craters ($\sim 0.25$).  The  average crater depth-to-diameter ratio is
$0.12 \pm  0.05$ \citep{bes10}. Small  craters have typically  a lower
depth-to-diameter  ratio.  These  characteristics are  consistent with
crater  degradation due  to  ejecta blanketing  and/or disturbance  of
loose  regolith on  the surface  triggered by  impact  seismic shaking
\citep{ric05},  in agreement with  results for  Itokawa \citep{hir09}.
Furthermore,  Ruby has  sharper  rims and  a higher  depth-to-diameter
ratio ($>0.14$; the bottom of the crater is not visible in the images)
than  several impact  craters.   Concerning the  size distribution  of
Steins'  surface material, the modeling  of the  observed strong
  opposition  effect and  the overall  photometric  properties suggest
  that Steins  may be  covered by  a layer of  regolith having  a mean
  particle  size  of  $\sim  100$~$\mu$m.   In  the  light  of  these
considerations, the observed large degree of crater degradation may be
explained in terms of  regolith blanketing, i.e.  deepest and sharpest
craters are younger than more degraded ones.\\

\section{The impactor flux}
 
\begin{figure*}[h]
\includegraphics[width=7cm,angle=-90]{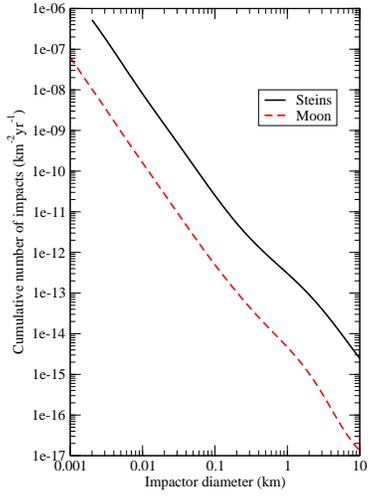}
\includegraphics[width=7cm,angle=-90]{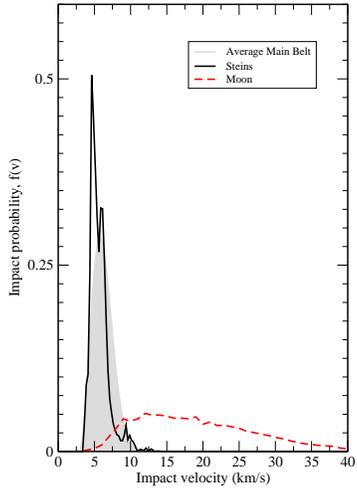}
\caption{Steins' cumulative  impactor size distribution  (upper panel)
  and impact  velocity distribution (lower panel) used  in the present
  work.  For a  comparison, the  lunar impact  distributions  are also
  shown \citep{mar09}.}
\label{impactors}
\end{figure*}

The  flux of impactors  can be  expressed in  terms of  a differential
distribution,  $\phi(d,v)$, which  represents the  number  of incoming
bodies per unit of impactor  size ($d$), impact velocity ($v$) and per
unit  time.  Such  a differential  distribution, under  the assumption
that   the   impact   velocity   is  independent   on   the   impactor
size\footnote{This  assumption stems from the  lack of systematic
    differences, at least for the available observational data, in the
    orbits of MBAs according to their sizes.}, can be written as:

\begin{equation}
\phi(d,v)=h(d)f(v) \label{phi}
\end{equation}

where  $h(d)$  is the  impactor  differential  size distribution,  and
$f(v)$ is the distribution of impact velocity (i.e. impact probability
per  unit impact  velocity)  normalized to  $\int  f(v)\,dv \equiv  1$
\citep[see][for details]{mar05,mar09}. In principle, the impactor size
distribution could  be derived from  observations.  At the  small size
range  relevant for Steins  ($d<1$~km), however,  there is  too little
observational  information to  enable such  an approach.   To overcome
such lack of information, we  use the average size distribution of the
main  belt  model  derived  by \citet{bot05a,bot05b}.   The  estimated
number  of impactors  at Steins  is then  obtained by  multiplying the
average size  distribution by  the intrinsic probability  of collision
with  Steins,  $P_i$ (see  fig.   \ref{impactors}  upper panel).   The
latter,  evaluated  by  taking   into  account  the  observed  orbital
distribution       of        main       belt       asteroids,       is
$P_i=2.87\cdot10^{-18}$~km$^{-2}$yr$^{-1}$,  nearly equal to  the main
belt average intrinsic probability  of collision.  The impact velocity
distribution  $f(v)$ for  Steins  has been  evaluated considering  the
population of  main belt asteroids that presently  intersect its orbit
(see fig.  \ref{impactors} lower  panel).Computations have been done
using the \cite{far92} algorithm. The resulting average impact
velocity is 5.7~km/s.\\

\section{The scaling law}

\begin{figure*}[h]
\includegraphics[width=10cm]{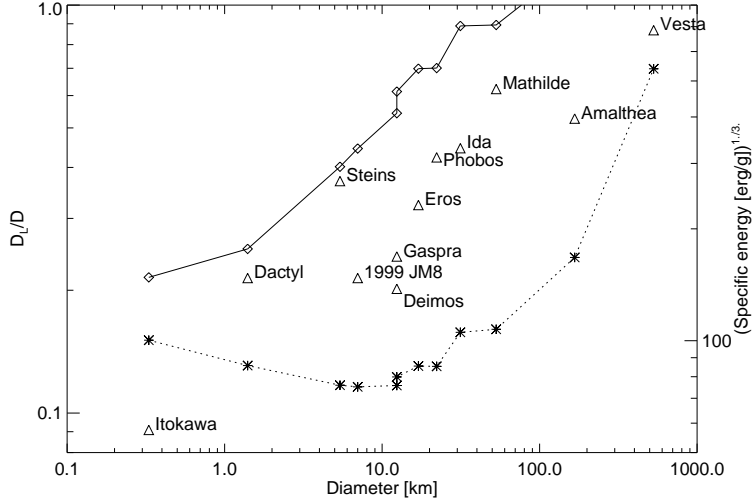}
\caption{The ratio of the  diameter of the biggest crater divided
    by the  average diameter  of various asteroids  (triangles, left
    ordinate) and the third root of the specific energy for disruption
    (diamonds) and shattering (asterisks)  of a basaltic body impacted
    with  a velocity of  5~km/s \citep[][right  ordinate]{ben99}.  The
    relation between  the left and  right ordinate is  arbitrary.  The
    critical impact is  expected to be between the  shattering and the
    disruption  limit.  For Steins  we used  an average  diameter of
    5.3~km, a size of the largest crater of 2~km and assumed a density
    (needed for the energy of  disruption) of 2~g/cm$^3$. Data for the
    other asteroids  are taken  from \citet{asp08}.  For  Amalthea and
    Vesta the specific  energy of disruption is above  the plot range.
    The figure shows that the ratio between size of the largest crater
    and body size  is not particularly high on  Steins. However, since
    the  energy  needed  to  shatter   a  body  is  at  a  minimum  at
    approximately  the size  of Steins,  an impactor  large  enough to
    create  the Ruby  crater  would  be expected  to  shatter (but  not
    necessarily   disrupt)   Steins.    The  model   calculations   by
    \citet{jut10} suggest  that this actually may have  been the case:
    In many cases  an impactor of the size needed  to create the Ruby
    crater will strongly damage  the original body, but Steins' global
    shape will remain  approximately the same and after  the event the
    crater will be clearly visible. }
\label{c_a}
\end{figure*}

\begin{figure*}[h]
\includegraphics[width=11cm,angle=-90]{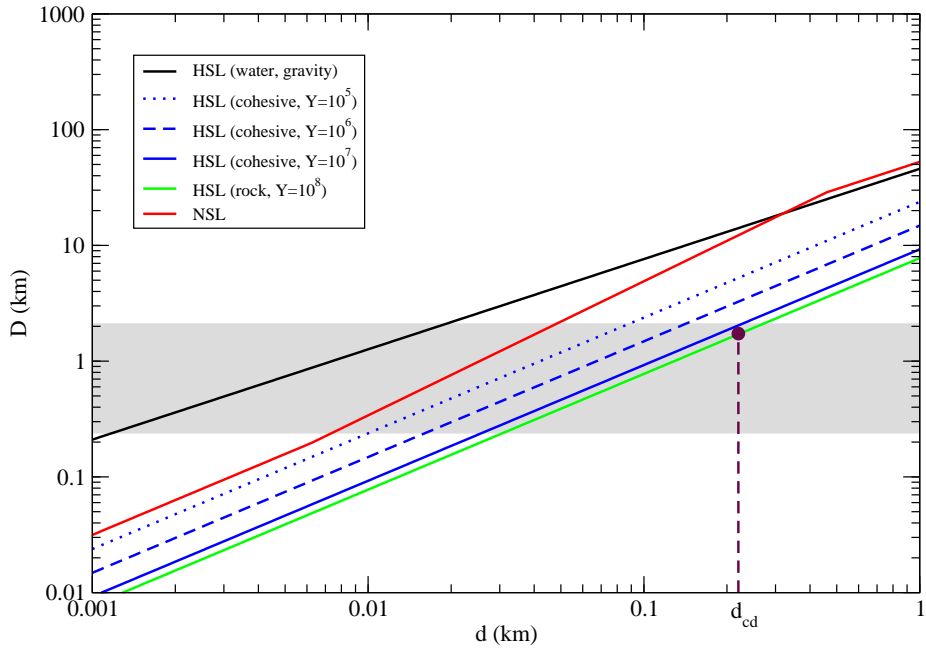}
\caption{Relationship  between  impactor  diameter  ($d$)  and  crater
  diameter ($D$) according to the scaling laws discussed in this work.
  We also indicated the approximate impactor diameter for catastrophic
  disrupt ($d_{cd}$)  of Steins, obtained for  an unfractured silicate
  body and HSL.  The limiting case  of the water scaling law (not used
  for dating purposes)  is shown only for a  comparison. The gray area
  indicates the crater size range detected on Steins.}
\label{sl}
\end{figure*}

The impactor  flux is converted into a  cumulative crater distribution
(the  so-called Model Production  Function, MPF)  using a  scaling law
(SL).  The physics  of the cratering processes for  small asteroids is
still poorly known.   In this work we use the  most updated SL derived
from  experimental analysis  \citep{hol07}  and hydrocode  simulations
\citep{nol96}.  The  SL for cratering on asteroids  depends on several
parameters,  the  most  important  being the  internal  structure  and
tensile strength  ($Y$).  Both are unknown for  Steins.  However, some
constraints can be obtained from morphological studies.\\
Let us first
consider  the  \citet{hol07}  scaling  laws  (HSL,  hereinafter).   An
interesting  issue   concerns  the  minimum   impactor  dimension  for
catastrophic disruption ($d_{cd}$) of Steins.  Assuming that Steins is
an unfractured  silicate rock and  using the relevant  specific energy
for   disruption,  $Q^{*}_{D}=1-2\cdot10^7$~erg/g   \citep{hol09},  we
derive that an impact at  an average modulus velocity of 5.7~km/s with
a  body  having size  $d_{cd}=0.20-0.25$~km  would  be sufficient  for
catastrophic  disruption.   Moreover,  for  an unfractured  rock  with
surface gravity  $g$ and density $\rho$, the  transition from strength
to gravity  cratering occurs at a crater  diameter of $\sim0.8Y/g\rho$
\citep{asp96}, which exceeds  the size of the Ruby  crater, except the
cases  of unreasonably  low $Y$  values for  a silicate  body (namely
$<10^5$~erg/g).   Therefore, using  HSL the  strength  regime applies.
Under these conditions, we obtain that a 2~km crater is produced by an
impactor having $d\sim d_{cd}$. \\
 In this respect it is interesting to note that even though the visual
 appearance of Steins  is dominated by the big  Ruby crater, the ratio
 of largest  crater diameter to  average asteroid diameter of  0.38 is
 not particularly  high.  Asteroids  Ida (0.44), Mathilde  (0.62), and
 Vesta   (0.87)  reach   considerably  larger   values  \citep{asp08}.
 However, if compared  to the specific energy required  to disrupt the
 body, the  big crater of  Steins stands out (Fig.   \ref{c_a}).  
   Therefore  the existence  of the  Ruby  crater may  be an  indirect
   evidence  that Steins  was not  a solid  rock at  the time  of Ruby
   formation.   Even if  a particular  tuning of  the  parameters may
 leave open  such a possibility, it  is more likely that  Steins was a
 rubble  pile or a  collection of  cohesive rubble  of rocks.   In the
 first case,  for a  pure cohesiveless rubble  pile the  gravity scale
 would apply.  In the latter case,  a cohesive rock scaling law may be
 more suitable.   Concerning the  present state of Steins,  it is
   likely  a rubble  pile  independently  of its  state  prior to  the
   formation  of  Ruby  \citep{jut10}.   This conclusion  is  also  in
   agreement  with the two  large fracture-like  features seen  on its
   surface  (see  previous  sections).  A preliminary  study  of  the
 formation of  Ruby crater has  been recently performed  via numerical
 modeling \citep{jut10}.  It has  been found that  a rubble  pile body
 with  some micro-porosity would  have survived  the formation  of the
 Ruby  crater,  although  the  non-porous monolithic  body  hypothesis
 cannot be ruled out.\\
 A further indication  in favor of the rubble  pile (both cohesiveless
 or with some low cohesion) nature  of Steins may come from its shape,
 possibly due to YORP spin-up \citep{kel10}. \\
In   conclusion,   from  previous   reasoning,   we     limit   our
investigations  to HSL  for cohesive  soils, and  test the  effects of
different  tensile strength.   As a  limiting case  of  a strengthless
material  we use  the HSL  for water  \citep{hol07}.   These equations
read:\\
 
\begin{equation}
D=kd\biggl(\frac{Y}{\rho v_{\perp}^2}\biggr)^\frac{\mu}{2}\biggl(\frac{\rho}{\delta}\biggr)^{\nu} \label{hh1}
\end{equation}
\begin{equation}
D=kd\biggl(\frac{gd}{2v_{\perp}^2}\biggr)^{-\frac{\mu}{2+\mu}}\biggl(\frac{\rho}{\delta}\biggr)^{-\frac{2\nu}{2+\mu}} \label{hh2}
\end{equation}

respectively for cohesive soils and water. $D$ is the crater diameter,
$v_{\perp}$ is  the normal component of the  impact velocity, $\delta$
is the  impactor density, $k, \mu,  \nu$ depend on  the material and
are  derived from  experiments.  Their  numerical values  are $k=1.03,
1.17,  \mu=0.41,  0.55,$  respectively  for  Eq.   \ref{hh1}  and  Eq.
\ref{hh2}, while  $\nu=0.4$ in all cases \citep{hol07}.
Concerning  the  strength,  we   may  regard  typical  lunar  regolith
($Y\sim10^5$~dyne/cm$^2$) and bulk silicates ($Y\sim10^8$~dyne/cm$^2$)
as limiting cases for a silicate body.
Highly    under-dense    (porous),    aggregate    materials    having
$Y<10^5$~dyne/cm$^2$  can  be  ruled  out  because  of  Steins'  stony
composition (E-type).   Concerning the higher limit,  a more realistic
estimate of the  strength for an asteroid may  be obtained considering
that the strength depends on the asteroid size $R$, with larger bodies
being weaker than smaller  ones of similar composition.  Assuming that
the  strength scales  as $R^{-1/3}$  \citep{asp96}, Steins'  hard rock
strength may be as low as a few $Y\sim10^6$~dyne/cm$^2$.  In the light
of previous reasoning,  we restrict the following analysis  to the HSL
for cohesive soils (for two representative values of tensile strength,
namely $Y=10^5,10^6$~dyne/cm$^2$).  In fig.  \ref{sl} the HSL obtained
for different parameters are reported.\\
A different  approach is  that proposed by  \citet{nol96}. They
estimated the cratering scaling law (NSL, hereinafter) using hydrocode
simulations.   Their main  result is  the discovery  of  the so-called
fracture regime,  which occurs in  between the two  extreme situations
represented  by the  strength  and gravity  regimes. Basically,  small
craters  are  formed in  the  classical  way,  with their  size  being
controlled  by the  local strength.   In large  craters, on  the other
hand,  the shock  wave propagates  ahead of  the excavation  flow, and
therefore the material is totally  fractured prior to its removal.  If
the  amount of excavated  material is  large enough,  the size  of the
resulting crater is controlled by the gravity.  A NSL has been derived
for  Gaspra,  which can  be  rescaled  to  other asteroids  using  the
approach  proposed  by \citet{obr06}.  Note  that  in  this
scenario bodies can survive much larger impacts than predicted by HSL.
NSL can be arranged in the following manner:

\begin{equation}
D=cd^{\alpha}\biggl(\frac{v}{v_0}\biggr)^{\beta}\biggl(\frac{g}{g_0}\biggr)^{\delta} \label{nsl1}
\end{equation}

where $c$,  $\alpha$, $\beta$ and  $\delta$ have different  values for
the strength,  fracture and gravity regimes.   The transitions between
strength  to fracture  regime and  from fracture  to  gravity regime
occur respectively at:

\begin{equation}
d_{sf}=d_0\biggl(\frac{v}{v_0}\biggr)^{\gamma} \label{nsl2}
\end{equation}
\begin{equation}
d_{fg}=d_1\biggl(\frac{v}{v_0}\biggr)^{\phi}\biggl(\frac{g}{g_0}\biggr)^{\theta} \label{nsl3}
\end{equation}

where  $d_0$, $v_0$  and  $g_0$ are  parameters  computed for  Gaspra.
Numerical   coefficient  used  here   (expressed  in   c.g.s.   units)
are\footnote{Notice   that  the   parameters  involved   in  equations
  \ref{nsl1},\ref{nsl2},\ref{nsl3} are not totally independent because
  of  continuity  conditions  at  the boundary  between  two  adjacent
  regimes.    They   can   be    written   in   the   following   way:
  $\beta=2\xi/(1-\xi),  2\alpha_f\xi/(1-\xi),  2\xi;  \delta=0,0,-\xi$
  for   the  strength,  fracture   and  gravity   regimes.   Moreover,
  $\gamma=-2\xi/(1-\xi);
  \phi=2\xi(1-\xi-\alpha_f)/[(1-\xi)(\alpha_f-\alpha_g)];
  \theta=-\xi(1/(\alpha_f-\alpha_g))$,  where the  subscript  f and  g
  stands for the fracture and  gravity regime and $\xi$ is a parameter
  that depends on the material and has been set to 0.22 \citep{obr06}.
  Note   that  the   above   equation  have   been  simplified   using
  $\alpha_s=1$.}:   $c=35,  26.61,   161.4;  \alpha=1,   1.159,  0.78;
\beta=0.56,  0.65,  0.44; \delta=0,  0,  -0.22$  respectively for  the
strength,  fracture  and  gravity regimes,  while  $\gamma=\phi=-0.56;
\theta=-0.58$     for    all     cases.      Finally,    $d_0=560$~cm,
$d_1=2.56\cdot10^4$~cm,  $v_0=5.0$~km/s  $g_0=0.448$~cm/s$^2$.  Figure
\ref{sl} shows the NSL rescaled to Steins.\\
By comparing  the SLs  reported in fig.   \ref{sl}, a large  degree of
variation  emerges.  For  a  fixed impactor  size  $d$, the  resulting
crater diameter $D$ may vary by  more than a factor of 10. However, if
we restrict ourselves to the  observed crater size range on Steins and
to    the   most   likely    scaling   laws    (NSL   and    HSL   for
$10^5<Y<10^6$~dyne/cm$^2$) the variation is within a factor of 3. 
  The  difference can be  partly explained  by the  fact  that NSL
  overestimates  cratering  efficiency  since  it neglects  the  shear
  resistance of materials \citep{nol96}. 
%

\section{The model production function}

Using  the  considerations  described  in  previous  sections,  it  is
possible to  compute the differential distribution  ($\Phi(D)$) of the
number of craters  with respect to their diameters  expressed per unit
time and surface area.  The MPF can be obtained by:

\begin{equation}
{\rm MPF}(D)=\int_{D}^\infty \Phi(\tilde{D})d\tilde{D}
\end{equation}

\begin{figure*}[h]
\includegraphics[width=10cm,angle=-90]{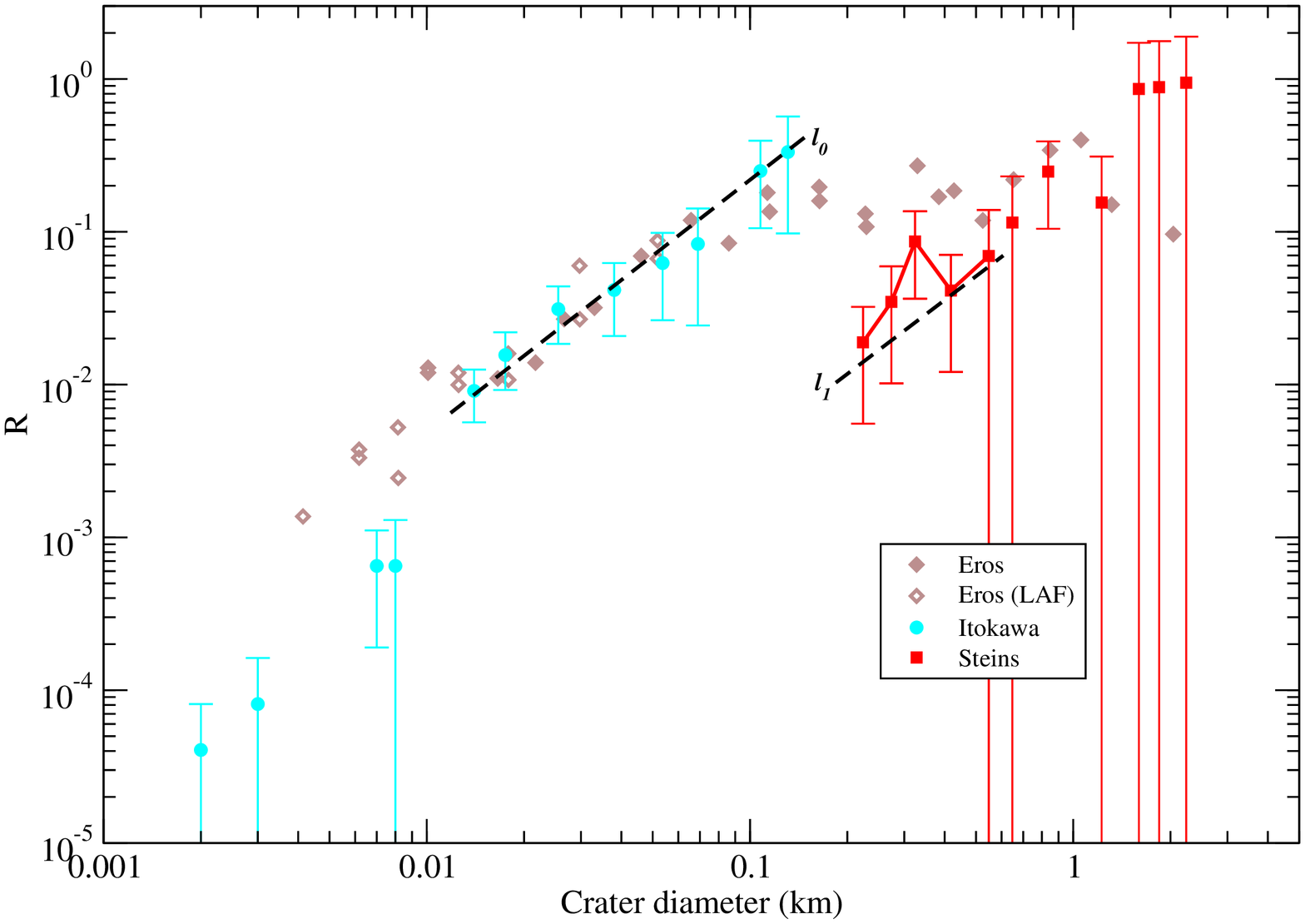}
\caption{Steins'  crater  data  shown  on a  Relative  plot  (R-plot).
  R-values  are computed according  to \cite{cra79}.   For comparison,
  Itokawa  and Eros  data  are  also shown.   Itokawa  data have  been
  computed  from table~1  of \cite{hir09},  while Eros  data  are from
  \cite{cha02}.  Dashed  line $l_o$  indicates the average  slopes for
  the left-most part  of the Itokawa/Eros distributions ($D<0.15$~km),
  showing  a paucity  of small  craters.   For Itokawa  and Eros  this
  paucity  of small  craters has  been  interpreted as  the result  of
  seismic  shaking \citep{ric04,mic09}.  Steins  also shows  a similar
  behavior  (see  line $l_1$  which  is  parallel  to $l_o$)  but  for
  $D<0.35$~km  the   trend  seems  to   have  a  steeper   slope  than
  Itokawa/Eros, possibly due to  different processes of crater erasing
  (see text for details).}
\label{rplot}
\end{figure*}

The distribution of craters for a given age $t$ is simply obtained by:
${\rm MPF}(D)\times  t$.  These  equations implicitly assume  that all
craters  accumulate  over  time  without interfering  with  previously
formed  craters (i.e.  no crater  erasing) and  that the  flux is
  constant  over  time.  The  latter  assumption,  according to  lunar
  chronology,   is   valid   for   ages   less   than   $\sim3.7$~Ga
\citep[e.g.][]{mar09}.  In  figure \ref{steins} (upper  panel) the fit
of the MPF to Steins crater counting data is shown.\\
The  first  important result  is  that  the  shape of  the  cumulative
distribution  cannot  be satisfyingly  fitted  in  the available  size
range.  In particular, when  fitting craters larger than $\sim0.6$~km,
the  smaller  craters  are  strongly underrepresented.   Similar  {\it
  kinks} in the cumulative  distribution have been observed elsewhere,
in particular on the  Moon and Mars \citep[e.g.][]{hie02,har09}, and usually
are attributed to episodes of  crater erasing which are more effective
for small diameters.   A similar lack of small  craters, but for sizes
$<10$~m, has also recently been observed on Itokawa \citep{hir09,mic09}.
In  fig.  \ref{steins}  (upper panel)  the best  fit for  large crater
diameters is shown.   The corresponding age, derived by  using the NSL
is $0.1  \pm 0.02$~Ga.  This estimate  is likely to  represent a lower
limit, since the NSL neglects  shear resistance and therefore tends to
overestimate  the crater  sizes  \citep{nol96}.  The  HSL derived  age
ranges   from  $0.28   \pm  0.06$~Ga   to  $0.94   \pm   0.2$~Ga,  for
$Y=10^5,10^6$~dyne/cm$^2$,  respectively.  Model  ages  are determined
through a $\chi^2$ fitting.  Formal errors are estimated considering a
variation of $\pm 30\%$ around the minimum $\chi^2$.
The cratering process may be  more complicated than assumed so far, in
particular the MPF may vary over  time.  The main reason for such time
dependence is that craters may erase over time.  A number of processes
responsible for crater erasing  on small bodies have been indentified:
local and global jolting, cumulative seismic shaking and superposition
of  craters \citep{gre94,gre96,ric04}.   Such effects  can  be modeled
\citep{obr06} and  the resulting MPF  can be written in  the following
manner \citep{mar09}:

\begin{equation}
{\rm MPF}(D,t)=\int_{D}^\infty \Phi(\tilde{D},t){\cal E}(\tilde{D},t)\,d\tilde{D}
\end{equation}

where the function ${\cal E}(D,t)$ is the ratio of the final number of
craters,  erasing  included,  to   the  total  number  (i.e.   erasing
excluded).    The  mentioned  erasing   process  depends   on  several
parameters.  As for the regolith jolting and superposition, we use the
parameters  adopted in \citet{obr06}  for Gaspra.   Cumulative seismic
shaking has  not been applied  in our simulations  due to the  lack of
detailed information on regolith mobility on Steins.\\
Figure \ref{steins} (lower panel)  shows the effects of crater erasing
on Steins' age determination.  Also  in this case, even though the MPF
is  shallower, the  small craters  are  not accurately  fitted by  the
models.  Derived ages are now increased with respect to what was found
neglecting crater erasing.  Focusing again on $D>0.6$~km, we derive an
age of $0.154 \pm  0.035$~Ga for the NSL. In the case  of the HSL, the
age   becomes  $0.49   \pm   0.18$~Ga  and   $1.6   \pm  0.5$~Ga   for
$Y=10^5,10^6$~dyne/cm$^2$, respectively.\\
We  also  investigated   whether  or  not  the  observed  crater
  population is saturated, i.e. has reached an equilibrium point where
  new   craters  erase   old  ones   leaving  unchanged   the  overall
  distribution.   We  find that  the  crater  population  has not  yet
  reached saturation. To see this point, in figure \ref{steins} (lower
  panel) we overplot the MPF for 0.5~Ga (NSL) and 3.6~Ga (HSL).  These
  ages have been chosen in  order to have the corresponding MPFs above
  all the cumulative data points.   Both MPFs are clearly separated by
  the best-fit  curves, indicating that the saturation  is not reached
  yet, therefore age assessment is possible. \\
It is  interesting that also when  taking into account  the erasing of
craters by superposition and  regolith jolt, smaller craters are still
strongly underrepresented.  A number  of possible explanations for the
origin of this kink may  be invoked.  For instance, small craters
  may have been  erased by regolith displacement due  to the effect of
  cumulative seismic  shaking.  This process has  been demonstrated to
  explain  the deficit  of  small craters  on  Eros \citep{ric04}  and
  Itokawa \citep{mic09}.  In order to  address this issue, we show the
  Steins' crater  distribution on  a R-plot (see  figure \ref{rplot}).
  R-plots  are useful  in  order  to analyze  fine  details of  crater
  distributions.   Data from  Itokawa and  Eros are  also overplotted.
  The comparison among Itokawa,  Eros and Steins is interesting mainly
  for the purpose of the small craters' depletion, although, it is not
  so  straightforward   because  of   the  different  size   range  of
  craters. However, some interesting comments can still be drawn.\\
Itokawa  and  Eros  show  a  marked depletion  of  small  crater,  for
$D<0.1$~km.  This  fact has been  interpreted as the result  of crater
erasing triggered by cumulative  seismic shaking due to repeated small
impacts  \citep{ric04,mic09}.  This   process  is  able  to  reproduce
accurately the  observed trend on Itokawa and Eros (indicated for clarity
by  the line  $l_o$  in fig.  \ref{rplot}).\\
As  for Steins, the  crater distribution  for $D<0.5-0.6$~km  shows an
overall  similar   behavior,  indicating  that  also   on  Steins  the
cumulative seismic shaking may be a viable explanation for the lack of
small craters.   Nevertheless there are some fine  details that differ
from Itokawa and  Eros which are worth to be  analyzed. To better show
these discrepancies, in fig.  \ref{rplot} we draw the line $l_1$ which
is parallel  to $l_0$. Despite the  large error bars,  the 3 left-most
bins ($D<0.3-0.4$~km) exhibit a steeper trend than that of Itokawa and
Eros, although  they might be  compatible due to the  relatively large
statistical errors.   This may  be an indication  that the  erasing on
Steins had a different origin, or at least that the cumulative shaking
is not the  unique responsible of the observed  lack of small craters.
We suggest that the Steins observed peculiar crater distribution could
retain the footprint  of an intense episode of  erasing triggered be a
single event,  likely the formation  of Ruby, which would  have erased
preferentially  craters below $D\sim0.3-0.4$~km.   Note that  the size
bin corresponding to the Ruby crater in the cumulative distribution is
close to the best fit model curve (see fig. \ref{steins}).  This means
that the formation  of such large craters is  already accounted for by
the model  and yet this is  not sufficient for explaining  the lack of
small  craters, at  least  with  the parameters  used  in this  study.
Therefore, we may argue that the formation of the Ruby crater happened
in a more recent time than the best fit age.  \\
An alternative explanation is connected to the YORP evolution that may
have triggered regolith mobility  and efficiently erased small craters
\citep[e.g. see][]{sch07}. \\
Without detailed  knowledge of the  internal nature of Steins  and its
regolith thickness  it is  very difficult to  draw a  firm conclusion.
Nevertheless, erasing  triggered by the Ruby impact  seems more likely
than  the one induced  by the  YORP effect.  This conclusion  seems at
least partially confirmed by the  fresh appearance of the Ruby crater,
with its sharp rims and high depth-to-diameter ratio.\\
  In any  case, we  may use the  crater distribution to  constrain the
  epoch of the putative  ``impulsive erasing''.  In order to correctly
  achieve this result  a detailed knowledge of the  erasing process is
  required.   This is  however far  beyond the  scopes of  the present
  paper, hence  we limit our discussion considering  two cases.  First
  of all, the  case where all small craters  ($D<0.35$~km) were erased
  at the  same time.  Through  MPF fitting of the  observed cumulative
  distribution for  small craters,  the approximate time  lapsed since
  the  erasing  can be  derived.   Considering  only  the craters  for
  $D<0.35$~km, for NSL we obtain an age of $0.032 \pm 0.004$~Ga, which
  becomes $0.072  \pm 0.01$~Gy  and $0.237 \pm  0.03$~Gy, for  HSL and
  $Y=10^5,10^6$~dyne/cm$^2$,  respectively  (see  fig.   \ref{steins},
  lower  panel).  Note,  however,  that the  accumulations of  craters
  during  this time  would  have  produced a  steeper  slope than  the
  observed   one   resulting  from   face   value   of  data   points.
  Nevertheless,  the  statistical errors  are  too  large to  conclude
  whether or  not the  model distribution is  really different  to the
  observed  trend.  In  this  respect,  it cannot  be  ruled out  that
  seismic shaking might  be entirely responsible for the  slope of the
  crater  size distribution  in  that  size range,  in  this case  the
  observed trend cannot  be used to constrain the  timescale of Ruby's
  formation.\\
A lower  limit for the epoch  of the impulsive erasing  can be derived
considering  that  it erased  most  or  all  of the  smallest  craters
($D\sim0.2-0.3$~km), but only  some of the larger craters  in order to
reproduce the  observed slope.  In  this case, the timescale  could be
set  by the  timescale to  form just  the smallest  craters.  The time
required to  accumulate the  4 observed craters  having $D=0.2-0.3$~km
spans from $\sim2$~Ma to $\sim10$~Ma, for NSL and HSLs respectively.\\
It  is possible,  using Poisson  statistics, to  compute  the expected
probabilities  that  the  impulsive  erasing  due  to  Ruby  formation
occurred  at the  computed  times.  Assuming  that  during the  crater
retention ages  the formation of only  a single crater of  the size of
Ruby  occurred, the  probabilities  that Ruby  event  occurred in  the
estimated recent times are: from  $\sim$1 to $\sim$17\% (NSL) and from
$\sim$0.6  to $\sim$12\%  (HSL),  respectively for  the two  different
estimates of the impulsive event times reported above.\\
Another consequence of  the formation of the Ruby  crater would be the
mixing of  the regolith  layer, with subsequent  reset of  the optical
properties.  It is interesting that detailed investigations across the
surface  of  Steins  have  shown  little or  no  spectral  variability
\citep{ley10}.   This  is somehow  in  contrast  with other  asteroids
visited by spacecraft, all showing  a certain degree of alteration due
to space  weathering \citep{cha04}.  The lack  of spectral variability
across Steins can be explained  in different ways, although it must be
noticed  that space weathering  response on  E-type materials  has not
been investigated  in details yet.  A  possibilty is that  the lack of
spectral  variability  may be  an  indication  of Steins'  insensitive
response  to   space  weathering   alteration  due  to   its  specific
composition.    However,  as   demonstrated   by  \citet{laz06},   all
asteroidal spectral types are found to show some degree of alteration.
In this case, Steins would have  either a fresh or a totally saturated
surface.  The fresh surface scenario  seems more likely given the fact
that  there  is no  spectral  variation  in  proximity of  small  -and
therefore  relativey  young- craters.   This  conclusion  would be  in
agreement with the relatively young age estimated for the formation of
the Ruby crater.

\section{Conclusions}

\begin{figure*}[h]
\includegraphics[width=8cm,angle=-90]{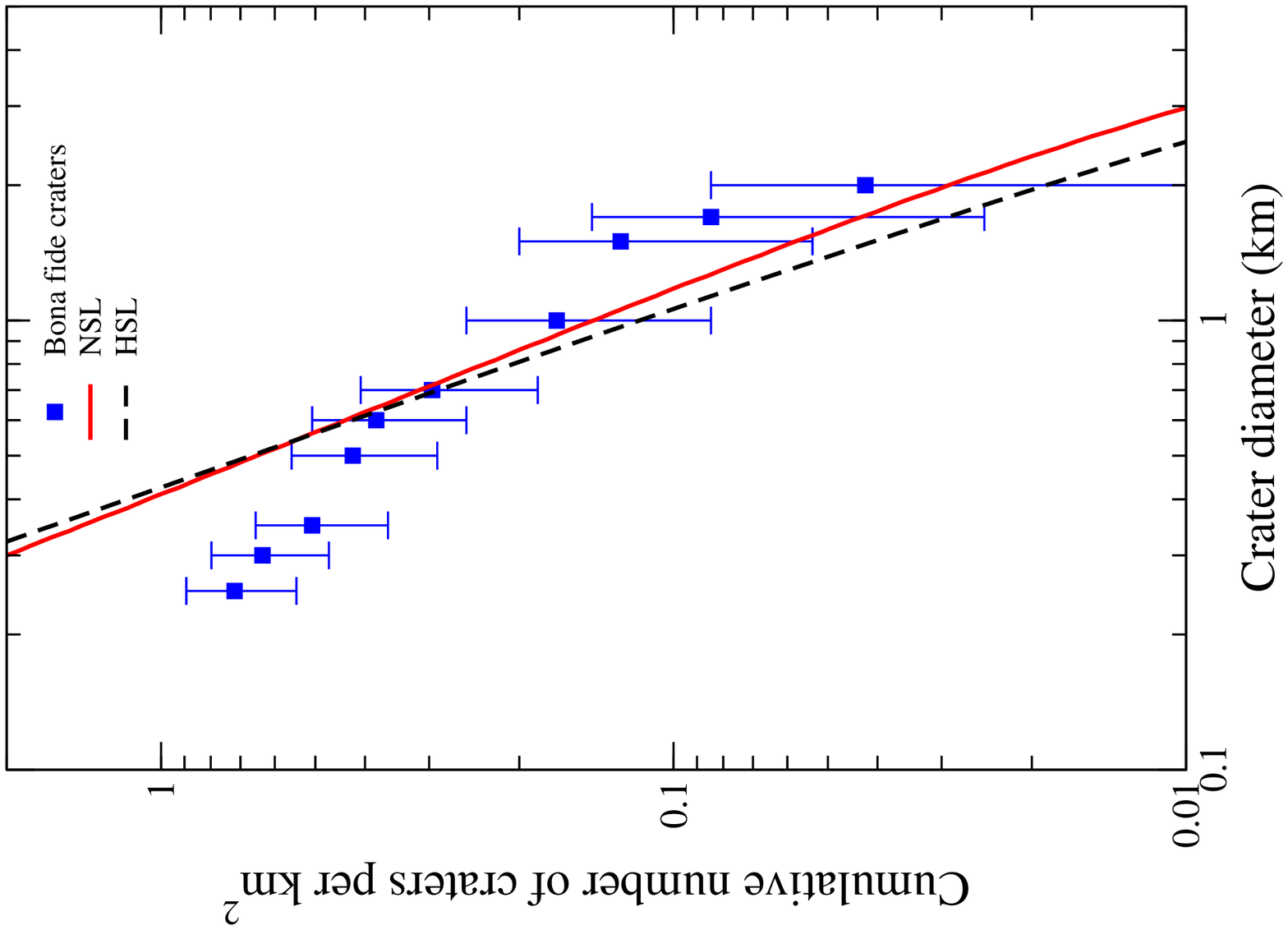}
\includegraphics[width=8cm,angle=-90]{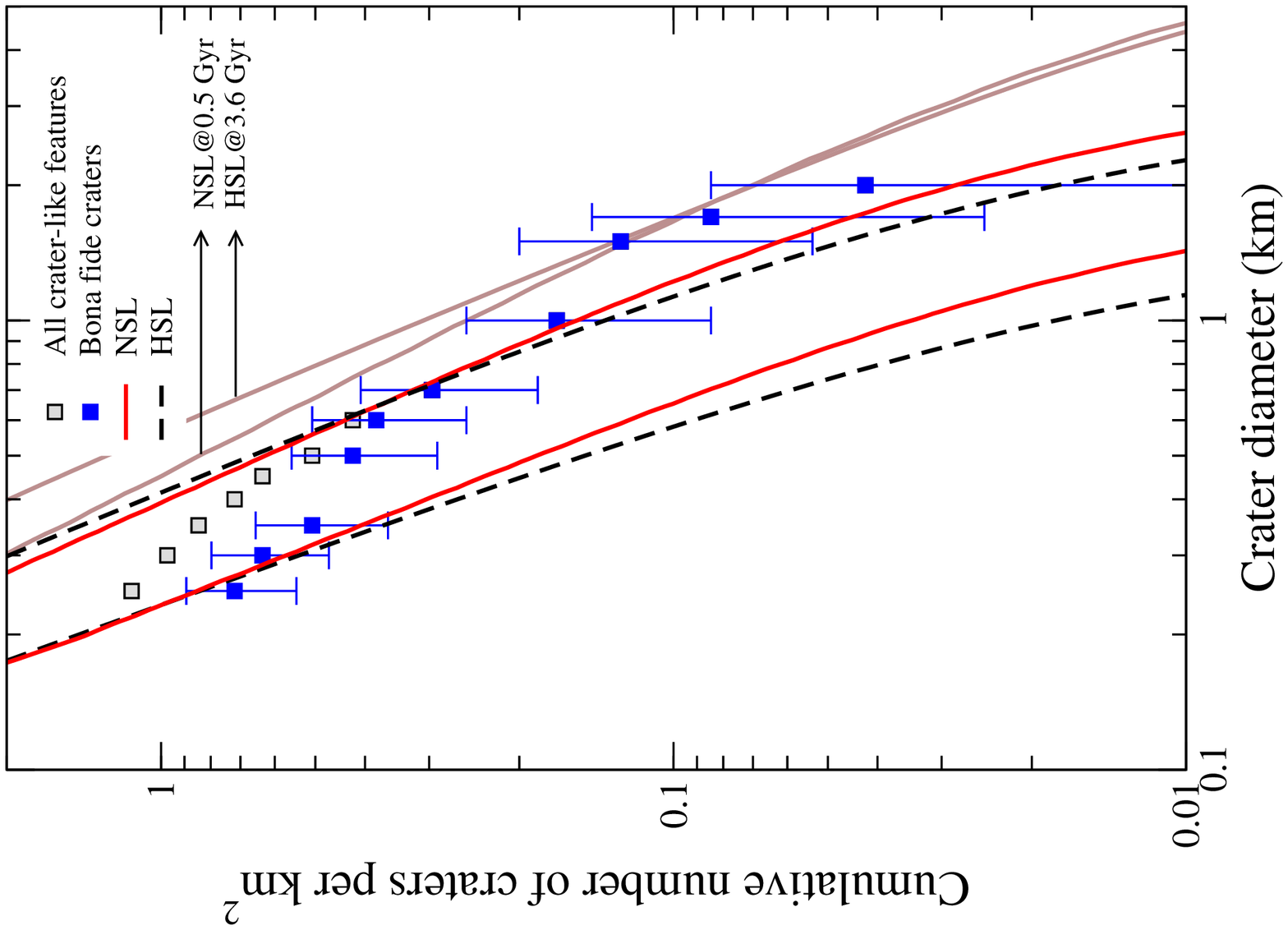}
\caption{Upper panel: Steins' age  estimates obtained with the MPF and
  no crater  erasing.  The best  fits have been  performed considering
  only  craters for $D>0.6$~km.   Notice the  strong lack  of observed
  craters for  $D<0.6$~km in comparison  to the models.   Lower panel:
  Steins' age estimates obtained with the MPF and crater erasing.  The
  best fits for small  diameters ($D<0.35$~km) is also shown, possibly
  indicating the  time of the formation  of the Ruby  crater (see text
  for  further details).  For a  comparison, the  distribution  of all
  crater-like features  is also shown  (errors bars are not  shown for
  simplicity,  see Fig.  \ref{allcraters}). To adreess
    the crater saturation  issue a couple of MPFs  for older ages have
    been also overplotted (see text).}
\label{steins}
\end{figure*}

In  this work  we  used crater  counting,  morphological analysis  and
impactor population modeling  to constrain Steins' cratering retention
age.  The derived ages vary according to the SLs used.  In particular,
using NSL and crater erasing  the derived age is $0.154 \pm 0.035$~Ga,
while using HSL and crater erasing the age ranges from 0.49 to 1.6~Ga,
according to the  values of strength used.  Moreover,  the modeling of
the crater erasing processes shows that the observed crater density is
not saturated  (at least for  the parameters adopted here).   The mean
collisional   age  of   Steins   is  estimated   to  be   $\sim2.2$~Ga
\citep[e.g.][]{mar06}.  Interestingly, similar  numbers apply also for
Gaspra.  Analogue  conclusions might be also valid  for the near-Earth
objects Eros, whose mean collisional age is $\sim1.7$~Ga \citep{mar06}
while cratering  age using NSL  gives 0.12~Ga \citep{obr06}  or, using
HSL, 1-2~Ga  \citep{mic09}.  The larger  bodies Ida and  Mathilde have
crater populations  either close to  the saturation or  saturated, and
consequently  their  cratering   age  estimate  is  less  constrained.
Despite the low number statistics,  the cratering age of the main belt
asteroids smaller than $\sim20$~km  seems to be systematically younger
than  their collisional  age.  This  result, if  confirmed  by further
studies, would  have important  implications on main  belt collisional
models. \\
Notably, the shape of Steins crater size distribution shows a kink for
diameters smaller than 0.5-0.6~km,  which may require a recent episode
of intense  erasing, although  seismic shaking could  have potentially
played a role in producing the observed distribution as well.  We also
attempt to constrain the epoch of such episode, possibly associated to
the  Ruby  crater  formation.   Focusing  on the  small  diameter  end
($D<0.35$~km) of the crater  cumulative distribution, and adopting the
MPF fitting we  obtain an age of $0.032 \pm  0.004$~Ga using NSL, from
0.072 to 0.237~Ga using HSL.   Under the assumption that the formation
of the  Ruby crater erased  all craters $<0.5-0.6$~km, the  above ages
could  possibly indicate  the time  since the  occurrence of  the Ruby
event.   A lower  limit  to the  age  of this  event,  can be  derived
considering the  time required to  accumulate the observed  craters in
the  range $0.2-0.3$~km.   This  produces an  age  from $\sim2$~Ma  to
$\sim10$~Ma, for NSL and HSLs respectively.\\
%
The derived ages  vary up to a  factor of ten depending on  the SL and
the tensile  strength used.  In  the present work we  investigated the
effects of a  relatively large range of SLs and  $Y$.  However, in the
light  of our  global understanding  of Stein  properties, we  favor a
crater retention  age ranging from $\sim0.15$ to  $\sim0.5$~Ga, and a
kink related event that could be as young as a few Ma up to
  some tens of Ma.\\
As a final remark, note that the conclusions derived in this paper are
based  on the  bona fide  crater distribution.   The  general scenario
outlined  (age, depletion  of  small craters)  remains  valid also  if
considering  all crater-like features  (see fig.   \ref{steins}, lower
panel). In  the latter case, however, the age  of the reset event
  is about a factor of two higher. \\

Acknowledgment

We thank  A.~Morbidelli for providing  us with the  impact probability
file for Steins.   We are also grateful to  the referee D.~O'Brien for
many  helpful comments  and for  providing  us the  published data  of
Itokawa and Eros cratering record.   Thanks also to a second anonymous
referee for insightful suggestions.   Finally, we thank E.~Simioni and
E.~Martellato for discussions.


\newpage

\end{document}